\documentclass[12pt]{article}
\pdfoutput=1

\usepackage{geometry}                		
\usepackage{physics}                   		 				
\usepackage{amssymb,bbm}
\usepackage{amsmath}
\usepackage{jheppub}

\newcommand{\nc}{\newcommand}
\nc{\La}{\Lambda}
\nc{\rnc}{\renewcommand }
\nc{\m}{\mu}
\nc{\f}{\varphi}
\nc{\n}{\nu}
\rnc{\k}{\kappa}
\rnc{\d}{\mathrm{d}}
\nc{\p}{\partial}
\rnc{\d}{\delta}
\nc{\g}{\gamma}
\rnc{\O}{\mathcal{O}}
\rnc{\t}{\tau}
\nc{\D}{\Delta}
\nc{\nn}{\nonumber}
\nc{\e}{\epsilon}
\nc{\w}{\omega}
\rnc{\a}{\alpha}
\rnc{\b}{\beta}
\rnc{\[}{\begin{equation}}
\rnc{\]}{\end{equation}}
\rnc{\(}{\left(}
\rnc{\)}{\right)}
\rnc{\l}{\lambda}
\rnc{\H}{\mathcal{H}}
\rnc{\r}{\rho}
\nc{\s}{\sigma}
\nc{\x}{\chi}
\rnc{\Lambda}{\kappa}
\nc{\<}{\left<}
\rnc{\>}{\right>}
\rnc{\|}{\Big\vert}

\title{Comments on the $S_N$ orbifold CFT in the large $N$-limit}

\author[]{Konstantinos Roumpedakis}

\affiliation[]{C. N. Yang Institute for Theoretical Physics\\
Stony Brook University\\
Stony Brook, NY 11794, USA. }

\emailAdd{konstantinos.roumpedakis@stonybrook.edu}

\abstract{
We elaborate on various aspects of the conformal field theory of the symmetric orbifold. We collect various results that have appeared in the literature, and we present a coherent picture of the operator content of this CFT, relying on the orbifold extension of the Virasoro algebra. We then focus on the large $N$-limit of this theory, discuss the OPE of two twist operators, and find various selection rules. We review how to calculate four-point functions of twist operators, and we write down the most general four-point function in the covering space for large $N$. We show that it depends on some functions that obey a set of algebraic equations, that resemble the scattering equations. Finally, we provide a recipe on how to calculate correlation functions with insertions of the orbifold Virasoro generators.
}

\preprint{YITP-SB-18-07}

\begin{document}

\maketitle

\section{Introduction}

Orbifold conformal field theories (CFTs) initially appeared in string theory \cite{Dixon:1985jw,Dixon:1986qv}, and have played a role in many related areas of theoretical physics ever since. A non-exhaustive list of examples includes instanton calculations \cite{Vafa:1994tf}, black hole physics \cite{Strominger:1996sh}, matrix string theory \cite{Dijkgraaf:1996xw,Motl:1997th,Dijkgraaf:1997vv,Dijkgraaf:1998zd}, and AdS/CFT  correspondence \cite{Maldacena:1997re,deBoer:1998kjm,Dijkgraaf:1998gf,Seiberg:1999xz,Larsen:1999uk,Lunin:2002fw,Gomis:2002qi,Gava:2002xb,Gaberdiel:2007vu,Dabholkar:2007ey,Pakman:2007hn,Taylor:2007hs,Giribet:2008yt}. Undoubtedly, on of the most exciting application of the orbifolds is in the context of the gauge/gravity duality \cite{Maldacena:1997re,Gubser:1998bc,Witten:1998qj,Aharony:1999ti}, with lots of recent activity  \cite{Gaberdiel:2014cha, Haehl:2014yla, Belin:2014fna,Gaberdiel:2015mra, Baggio:2015jxa,Belin:2015hwa,Jevicki:2015irq}. 

Type IIB string theory, compactified on $AdS_3 \times S^3 \times M$, with $M$ either $T^4$ or $K_3$, has a region in its moduli space which is thought to be dual to a deformation of the symmetric orbifold of $N$ copies of a supersymmetric sigma model with target space $M$. 
The purpose of this paper, given the importance of the $S_N$ orbifold in holography, is to elaborate on various aspects of the structure of the orbifold CFT.

Although the general structure of the different sectors of the theory is well-known, so far there are not been many attempts to explicitly construct its Virasoro primary operators. With that said, we describe a general recipe of how to construct these operators using an orbifold extension of the usual Virasoro algebra \cite{Borisov:1997nc}, and we study the insertions of these operators in correlation functions. Having this picture in mind, we study the anatomy of the operator product expansion (OPE) of two twist operators in the large $N$-limit, and we present a set of constraints that dictates which operators can appear. Moreover, we calculate the most general four-point function of twist operators in the covering space, up to some unknown functions that obey an algebraic set of equations, similar to the scattering equations \cite{Cachazo:2013iaa,Cachazo:2013gna}. Based on these results, one can calculate any four-point function of twist operators as a power series in the position of the fourth operator (we fix the position of the first three at $0$, $1$ and $\infty$). 

In section $2$, we review basic things about the $S_N$ orbifold. We start by considering $N$ copies of a sigma model with target space $M$, and for simplicity we focus on its bosonic sector, although extensions to fermions are straightforward \cite{Lunin:2001pw}. We define the orbifold theory and we review the general structure of its different sectors.

In section 3, starting from the fact that the operators of the sigma model have fractional modes in the twisted sectors \cite{Jevicki:1998bm}, we study the generating algebra of the orbifold \cite{Borisov:1997nc}, and we present a general picture of the operator content of the theory.

In section 4, we revisit the calculation of the four-point function of twist operators. These are notoriously difficult to compute \cite{Arutyunov:1997gt,Arutyunov:1998eq,Lunin:2000yv,Lunin:2001pw,Pakman:2009zz,Burrington:2012yq}. We fist consider the large $N$-limit where many simplifications can be made, and we determine which operators can appear in the OPE of two twist operators. Keeping in mind, that in this limit only the zero-genus covering space contributes, we explicitly construct the covering map from the base sphere to the covering sphere. After that, we give the general expression of the four-point function of four twist operators in the covering space.

In section 5, we study insertions of the orbifold algebra inside correlation functions, a technique that gives a way to calculate orbifold Virasoro blocks as a power series, like in the usual case of Virasoro blocks.

Finally, in section 6, we consider the simplest example of a four-point function and perform the comformal block decomposition for the first few blocks. We verify the above picture and we make contact with recent developments.

\section{Preliminaries}

We begin by reviewing some basic facts about the conformal field theory of the symmetric orbifold. Consider a two-dimensional CFT with central charge $c$. We will refer to this, as the {\it seed CFT}. We focus on the bosonic sector of the theory which is assumed to be a sigma model with action

\[S=2\int d^2 z \;  G_{ij} (X)\p X^i \bar{\p} X^j, \]
where $G_{ij}(X)$ is the metric of the target space $\mathcal{M}$, and $i=1,\dots D$ with $D$ the dimensions of $\mathcal{M}$.  Here we have taken the theory to live on the two-dimensional complex plane

\[ ds^2=dzd\bar{z}. \]

We define the orbifold CFT by considering $N$ copies of the seed CFT $X^i_I $ with $I=1,\dots N$, and we impose the equivalence relation

\[X^i_I \sim X^i_{g(I)}, \quad \forall  g \in S_N. \]
By $S_N$ here, we denote the group of permutations. This identification introduces new sectors, defined by the boundary conditions 

\[X^i_I( e^{2\pi i} z) = X^i_{g(I)}(z). \label{BC}\]

This is familiar from string theory, where in the case of the spinning string we can have two possible boundary conditions for the fermionic fields, namely Ramond and Neveu-Schwarz. In this case, we deal both sectors simultaneously by introducing spin fields. So, at the end we have one fermionic field which obeys trivial boundary conditions around every point, except from those points where spin fields are inserted. In other words, these spin fields create a Ramond vacuum when acting on the Neveu-Schwarz vacuum.

In the symmetric orbifold CFT, we introduce twist fields $\s_g(z)$ that create the twisted vacua. In the absence of twist fields, the $X_I$'s obey trivial boundary conditions, whereas in the presence of a twist field at position $z=z_1$, they obey 

\[ X^i_I( e^{2\pi i} (z-z_1))\s_g(z_1) = X^i_{g(I)}(z-z_1)\s_g(z_1). \]

Henceforth, we will omit the index $i$. Clearly, the twist operators $\s_g$ are not invariant under the action of $S_N$. We define ``gauge invariant" twist operators by

\[\s_{[g]}=\sum_{h \in S_N} \s_{h^{-1} g h}. \label{gauge invarinat sigma}\]
Hence, twist operators are in one-to-one correspondence with the conjugacy classes $[g]$, of the permutation group. We denote the resulting orbidold CFT by

\[Sym^N( \mathcal{M})\equiv \otimes^N \mathcal{M}/ S_N,\]
that corresponds to the symmetric product of $N$ copies of the theory with target space $\mathcal{M}$.

The Hilbert space\footnote{Here we restrict our attention to the subspace with zero momentum.} of the theory \cite{Dixon:1985jw,Dijkgraaf:1996xw} is the sum of all the sectors

\[ H\(Sym^N( \mathcal{M})\)= \bigoplus_{[g]} H_{[g]}. \label{Hilber space} \] 
The untwisted sector corresponds to the conjugacy class of the identity element. Recall that the conjugacy classes of $S_N$ have a specific cycle structure

\[[g]=(1)^{k_1} (2)^{k_2} \dots (N)^{k_N} , \quad \sum_{n} n k_n=N \label{class}.\]

 If a conjugacy class contains many cycles of the same size ($k_i>1$) we will have to further consider their symmetric product  
 
\[H_{[g]}= \bigotimes_n S^{k_n} H_{(n)}, \]
 where $H_{(n)}$ denotes the subspace created by a twist operator corresponding to a single-cycle of length $n$. These symmetric products are of the form
 
\[S^{k_n} H_{(n)}= \(H_{(n)} {\scriptstyle\bigotimes} H_{(n)} \dots {\scriptstyle\bigotimes} H_{(n)} \)^{S_{k_n} }.\]
Taking this into account, it is clear that the Hilbert space of the symmetric orbifold CFT can be fully understood once we have constructed the subspace $H_{(n)}$, which we will study in detail in the next section.

\section{Hilbert Space}

The aim of this section is to collect various aspects of the Hilbert space $H_{(n)}$ that can be found in the literature, and give a coherent prescription of how to construct it explicitly. Since this $H_{(n)}$ space is the same as in the case of $Z_n$ orbifold CFT \cite{Dixon:1986qv,Klemm:1990df,Borisov:1997nc}, we will do so by considering $n$ free scalars $X_I(z)$.

Because of the boudary conditions in (\ref{BC}), these scalar fields are not single-valued in the $z$ plane. For this reason, we replace this $n$ fields with a single field $X(z)$ that lives on a Riemann surface with multiple sheets. We call this the {\it covering space} \cite{Lunin:2000yv,Lunin:2001pw,Pakman:2009zz}. The $z$ coordinates cannot be globally defined on this space. Thus, in the covering space, we use another coordinate system $t$. The map $z(t)$ will be a holomorphic function with the right monodromies at the insertions of twist operators. For example, in the presence of a twist operator $\s_n(z_1)$, the covering space in the vicinity of $z_1$ will look like

\[z-z_1\approx (t-t_1)^n, \]
where $t_1$ the image of $z_1$. In the presence of only two twist operators, $\s_n(0)$ and  $\s_n(\infty )$, the covering map is just $z(t)=t^n$.

As we will discuss in section (\ref{sec:map}), in principle it is very hard to construct the covering map. Even if the base space is topologically trivial, the covering space may have non-trivial topology. The zero-genus maps for the case of three twist fields were explicitly constructed in \cite{Lunin:2000yv}. Whereas in the presence of four fields, as it was shown in \cite{Pakman:2009zz}, the problem of constructing the zero-genus maps can be reduced to solving some second-order differential equation, which unfortunately does not have a closed form general solution. We will revisit the construction of the covering map in section (\ref{sec:map}).

In the covering space the theory is just the usual free scalar CFT, and the holomorphic part of the Hilbert space can be generated by acting on the vacuum with the modes  $L_n^{cov}$ of the energy-momentum tensor $T(z)$, which obey the usual Virasoro algebra, and the first mode $\p X _1$ of the primary field $\p X(z)$. Throughout this paper, we focus on the zero momentum states and don't consider vertex operators, the reason being that this subspace is universal and does not depend on the topology of the target space. 

In the presence of the twist operator $\s_n(0)$,  the energy-momentum tensor will have a mode expansion 

\[ T_I (z)=\frac{1}{n}\sum_{m=-\infty}^{+\infty}  L_{m/n} z^{-\frac{m}{n}-2} e^{-2 \pi i \frac{m}{n} (I-1)} ,\label{T twisted}\] 
while for $\p X$ 
\[\p X_I(z) = \frac{1}{n}\sum_{m=-\infty}^{+\infty} \p X_{m/n} z^{-\frac{m}{n}- 1} e^{-2 \pi i \frac{m}{n} (I-1)},\label{X twisted} \]
where $I=1,\dots , n $. These mode expansions satisfy the boundary condition 

\[\p X_I( e^{2 \pi i} z)= \p X_{I+1}(z),\]
and similarly, for $T_I(z)$. In order to determine the algebra of these fractional generators, we consider two twist operators at $z=0,\infty$, with images at $t=0,\infty$ respectively. As we mentioned earlier, the covering map in this case is just $z(t)=t^n$. 

Inverting equation (\ref{T twisted}) for the modes we get

\[ L_{m/n}=\int \frac{dz}{2 \pi i} \sum_{I=1} ^n T^I(z)  e^{2 \pi i m (I-1)/n} z^{m/n+1} \label{L}.\]
When passing to the covering space, the energy-momentum tensor transforms as 
\[T(z)=\frac{1}{z'(t)^{2}} \(T^{cov}(t)-\frac{c}{12} \{z,t\}\), \label{T trans}\]
Here $T(z)=T^I(z)$ for $2\pi I<arg(z)< 2\pi (I-1)$. The function $\{z,t\}$ is the usual Schwarzian derivative

\[\{z,t\}=\frac{z'''(t)}{z'(t)}-\frac{3}{2}\frac{z''(t)^2}{z'(t)^2}=\frac{1-n^2}{2t^2}. \label{Schwartian}\]
Lifting expression (\ref{L}) to the covering space we get

\[L^{cov}_m=  n L_{\frac{m}{n}} - \d_{m,0}\frac{c}{24}  (n^2-1). \label{Lcov}\]
Imposing that the modes on the covering space $L^{cov}_m$ obey the usual Virasoro algebra, we obtain

\[ \left[ L_{\frac{m_1}{n}},L_{\frac{m_2}{n}} \right]= \frac{m_1-m_2}{n} L_{\frac{m_1+m_2}{n}}+ \frac{c}{12} \frac{m_1}{n} \left( \left(\frac{m_1}{n}\right)^2-1\right) \d_{m_1+m_2,0} \label{algebra}.\]
This algebra has appeared before in the literature \citep{Fuchs:1991vu, Borisov:1997nc}. We note that the argument can be reversed, and by demanding that these two types of generators satisfy the right algebras, one can constrain the covering map.

It is now trivial to determine the scaling dimension of the twisted vacuum, and consequently the scaling dimensions of the twist operator. The vacuum in the covering space is annihilated by $L^{cov}_0$ and hence equation (\ref{Lcov}) implies

\[L_0 \ket{n}= L_0 \s_n(0) \ket{0}=\frac{c}{24}  \frac{n^2-1}{n} \s_n(0) \ket{0}. \]
This gives the well-known formula for the scaling dimension of twist operators

\[h_n=\frac{c}{24}  \frac{n^2-1}{n}. \label{h_n}\]
Another observation is that since $L^{cov}_{-1}$ annihilates the vacuum in the covering space, then 

\[L_{-1/n} \s_n=0. \]

We are now ready to construct the Hilbert subspace, that corresponds to a single cycle of length $n$, by acting with $\p X_{-1/n}$ and $L_{-m/n}$ and their antiholomorphic counterparts, on the twisted vacuum. One can compare the resulting states above, with the partition on the torus \cite{Dijkgraaf:1996xw,Bantay:2000eq,Haehl:2014yla} 

\[ Z_N=\sum_{\{k_n\}} \prod_{n=1}^N \frac{1}{k_n!} (T_n Z(\t,\bar{\t}))^{k_n} ,\]
where $T_n$ is the Hecke operator defined by

\[  T_n Z(\t,\bar{\t})= \frac{1}{n} \sum_{d|n} \sum_{a=0}^{d-1} Z\( \frac{n \t + a d}{d^2},\frac{n \bar{\t} + a d}{d^2} \), \] 
and $Z(\tau,\bar{\tau})$ the partition function of a single scalar field. Expanding the above expression for $Z_N$ in a power series of $q=e^{i\tau}$ and $\bar{q}=e^{-i\bar{\tau}}$, one can verify that the multiplicities of states with the same dimension and spin match exactly.

We can now explore the operator content of the theory. The algebra in (\ref{algebra}) contains 
the usual Virasoro algebra as a subalgebra for $m$ multiples of $n$. This subalgebra is present in all sectors, and gives the modes of the full energy-momentum tensor

\[T(z)=\sum_{I=1}^N T_I(z).\]
The question that arises now is, which are the Virasoro primaries with respect to $T(z)$. In the untwisted sector, the Virasoro primaries consist of $S_N$-invariant sums of products of $\p X^I$, the first two being

 \[\p X=\sum_I \p X^I, \quad W_2= \sum_{I>J} \p X^I \p X^J. \label{gauge invariant operators}\]

In the twisted sectors things are more complicated. First of all, we can view the multi-cycle twist operators as ``normal-ordered" products of single cycles. In this case, ``normal-ordering" means cycles with no common elements. For instance,

\[ \s_{[2] [2]}= :\s_{[2]} \s_{[2]}:= \s_{(12)}\s_{(34)}+ \s_{(13)}\s_{(2 4)} +\dots \label{multicycles}\]
In other words, we just throw away terms like $\s_{(12)} \s_{(23)}$. These operators create the twisted vacua. We can further act with appropriate combinations of $\p X_I$'s or $T_I$'s to create excited twisted primary states. For instance,

\[ (\p X_1-\p X_2)(\p \bar{X}_1-\p \bar{X}_2) \s_{(12)}(z) \dots \label{xs}\]
creates a primary scalar state with scaling dimension $h=\bar{h}=h_2+\frac{1}{2}$ where $h_2$ is given by (\ref{h_n}). In terms of the modes in (\ref{X twisted}), this state is just $\p X_{-\frac{1}{2}}\p \bar{X}_{-\frac{1}{2}} \ket{\s_2}$. As we review in section (\ref{sec:large N}), the correlation functions of twist operators for large $N$, do not depend on the details of the target space. Therefore, operators like the ones in (\ref{xs}), do not contribute to leading order in $1/N$. Thus, in this limit, only primary operators built out of fractional generators are important. An example of such primary operators is

\[\t^k_n (z)=\sum_{I=1} ^n e^{ \frac{2 \pi i k (I-1)}{n}} T_I \s_n (z), \label{fractional s}\]
for $k=1, \dots, n-1$. The scaling dimension of this operator is $h=h_n+k/m$, and in terms of the fractional generators its highest weight state is $L_{-k/n}\ket{\s_n}$. All in all, we can build an infinite number of primaries under the integral Virasoro algebra, by acting on the vacuum with $L_{-m/n}$'s with $m<n$. In some cases, we might need to consider combinations of such terms in order for to make a primary, which has to be annihilated by all $L_m$ with $m>0$.

\section{Correlation Functions}

In this section we discuss correlation functions in the orbifold CFT. We begin by giving the path-integral definition following \cite{Lunin:2000yv,Burrington:2012yn}, and continue by briefly reviewing how one can evaluate it. Consider a correlation function of some gauge invariant operators

\[ \<\s_{[g_1]} (z_1)\s_{[g_2]} (z_2) \dots O_1(w_1)O_2(w_2)\dots  \>, \label{general corelation}\]
where the $O_i$'s operators in the untwisted sector. One can also consider excited twist operators, but we will come back to that in section \ref{sec:Ward}. Another possibility is to consider operators like the ones in (\ref{xs}), but we will not consider this case here \cite{Burrington:2012yn}. 
From (\ref{gauge invarinat sigma}), we see that a twist operator is a sum of gauge non-invariant terms $\s_n$. Likewise, the $O_i$'s are sums of non-invariant operators $o_i$, and in order to compute the above quantity, we have evaluate all terms like

\[ \<\s_{g_1} (z_1)\s_{g_2} (z_2) \dots o_1(w_1)o_2(w_2)\dots  \>, \quad g_i\in S_N.\]
A necessary condition for such a term to be non-zero is 

\[ g_1 g_2 \dots=1. \]

In general, some of the $g_i$'s can be multi-cycle elements of $S_N$.  As we discussed in (\ref{multicycles}), multi-cycle twist operators can be thought of as normal-ordered products of single-cycle operators. Hence, the evaluation of (\ref{general corelation}) boils down to the calculation of

\[ \<\s_{n_1} (z_1)\s_{n_2} (z_2) \dots o_1(w_1)o_2(w_2)\dots  \>, \]
where $\s_{n_1}$ are single-cycle twist operators and the $o_i$'s are gauge non-invariant operators in $O_i$. These terms can be calculated in principle from the path integral 

\[ \<\s_{n_1}(z_1) \s_{n_2}(z_2) \dots o_1(w_1)o_2(w_2)\dots  \>= \frac{1}{Z_1 \!^N} \int_{twisted} [DX] e^{-S} o_1(w_1)o_2(w_2)\dots  \nn\]
where $Z_1$ is the partition function of a single scalar field on the sphere. On the right-hand side, we have to evaluate the path integral with the appropriate boundary conditions at the positions of the twist operators. The strategy to evaluate this path-integral is to lift it in the covering space where the complication of the boundary conditions is absent. However, the map induces a metric in the covering space

\[ ds^2=dz d\bar{z} = \| \frac{dz}{dt} \|^2 dt d\bar{t}.\]
This metric can be rescaled back to one, at the expense of getting en extra factor because of the Weyl anomaly which takes into account the change of the measure. After doing so, we get

\[ e^{S_L}  \<z'(t_1)^{-h_1}o_1 (t_1) z'(t_2)^{-h_2} o_2(t_2)\dots\>,\label{cor function in cov space}\]  
where $S_L$ is the Liouville action \cite{Friedan,Polyakov:1981rd}, and it is given by

\[S_L= \frac{c}{96 \pi}\int d^2 t \sqrt{-g} \(\p_\m \phi \p^\m \phi + R \phi \), \quad \phi=\log\|{\frac{dz}{dt}}\|^2. \label{Liouville}\]

For more details, as well as for the case where the twist operators are dressed with $\p X_I$'s, see \cite{Lunin:2000yv,Burrington:2012yn}. We see that the calculation of correlation functions in the orbifold theory is reduced to the calculation of the Liouville action. However, as it was shown in \cite{Lunin:2000yv}, it requires regularization, and in the same paper, a precise way to regularize and evaluate it was given. 

In the rest of the paper we focus on correlation functions of twist operators only

\[ \<\s_{n_1}(z_1) \s_{n_2}(z_2)\dots  \s_{n_s}(z_s)\>. \label{4p twist}\]

In order to compute this correlation function, we need the explicit map to the covering space $z(t)$. This will be the subject of section (\ref{sec:map}). We conclude this section by reviewing how to organize all these terms as in \cite{Pakman:2009zz}.

 Terms like the ones in (\ref{4p twist}), can be organized into two different ways. First, by the number of participating copies $\x$ in the twist, which from the Riemann-Hurwitz formula determines genus of the covering space

\[g=\frac{1}{2} \sum_{j=1} ^s (n_j-1)-\x+1.\label{g}\]

Secondly, after computing the Liouville action (\ref{Liouville}), in order to express the result in the $z$ coordinates, we need to invert the map $z(t)$. This inversion gives many solutions and we need to sum over all of them. It was shown in \cite{Pakman:2009zz}, that all these solutions are in one-to-one correspondence with different distributions of the copies in the $\s$'s. For instance, the cases $\< \s_{(432)} \s_{(23)}\s_{(412)}\s_{(12)}   \>$ and $\< \s_{(142)} \s_{(23)}\s_{(342)}\s_{(12)}   \>$ correspond to different functions $t(z)$.
In summary, the connected piece will look like

\[ \<\s_{[n_1]}(z_1) \dots \s_{[n_s]}(z_s)\>_{conn}= \sum_g ^{g_{max}} \sum_{\a_g} C_{g,\a_g} \<\s_{g_1^{\a_g}}(z_1) \dots  \s_{g_s^{\a_g}}(z_s)\>_g ,\]
where on the left-hand side we have two sums, one for the covering spaces with different genera, and one for the inequivalent terms with the same genus. By $g_j^{\a_g}$ we denote the cycles of length $n_j$ which give a covering space of genus $g$, and by $C_{g,\a_g}$ the number of terms with the above specifications. We will comment on these constant in the next section. The above formula may seem very complicated for practical purposes, but we will see that great simplifications happen in the large $N$- limit.

\subsection{Large N-Limit \label{sec:large N}}

In this section we review the $N$-dependence of the correlation functions in (\ref{4p twist}) following \cite{Jevicki:1998bm,Lunin:2000yv,Pakman:2009zz}. The goal is to illuminate the structure of the OPE of twist operators in the large $N$-limit. We also argue that in this limit, this kind of correlation functions does not depend on the details of the target space, a property known as {\it universality} in the literature \cite{Lunin:2000yv}.

As we already mentioned, the two-point function of gauge invariant operators (\ref{gauge invarinat sigma}) splits into many gauge non-invariant terms. Summing over all these contributions introduces an $N$-dependent factor, which can be absorbed by appropriately normalizing the $\s$'s. A simple counting argument, shows that the two-point function of the unnormalized operators is

\[ \< \s_{[n]} (0) \s_{[m]} (z)\> = n N! (N-n)! \frac{\d_{n m}}{|z|^{2h_n}}. \]
Note that there are $N!/(N-n)!$ ways to choose $n$ out of $N$ copies to twist. Moreover, we can further commute the $N-n$ non-participating copies in all possible ways, contributing a factor of $(N-n)!^2$. We further get an additional factor of $n$ which accounts for the $n$ possible cyclic permutations of the twisted copies. Hence, we normalize these operators by defining

\[\hat{\s}_{[n]}= \frac{1}{\sqrt{n N! (N-n)!}} \s_{[n]}. \]
Similarly, as in \cite{Pakman:2009zz}, for the $s-$point function we have

\[ \<\hat{\s}_{[n_1]}(z_1) \dots  \hat{\s}_{[n_s]}(z_s)\>_{conn}= \sum_g \sum_{\a_g} \hat{C}_{g,\a_g} \<\s_{g_1^{\a_g}}(z_1) \dots  \s_{g_s^{\a_g}}(z_s)\>_g ,\]
where

\[\hat{C}_{g,\a_g}=\frac{N!}{(N-\x)!}\prod_j ^s \frac{\sqrt{n_j (N-n_j)}}{N!} .\]
The large $N$-limit of these coefficients, by use Stirling's formula, is

\[\hat{C}_{g,\a_g} \sim N^{1-g-\frac{s}{2}}. \label{C coeff}\]

Amazingly enough, we see that to leading order in $N$, only the $g=0$ terms contribute, which corresponds to the covering space being a sphere. Another simplification in this limit, is that the correlation functions of twist operators are universal, and do not depend on the details of the target space $\mathcal{M}$. 
From the above discussion, we see that a general correlation function will have an expansion like

\[ \<\s_{[n_1]}(z_1) \dots  \s_{[n_s]}(z_s)\>_{conn} =\frac{1}{(Z_1)^\x} \( F^{(s)}_0(z_i) Z^{cov}_{g=0}+ \frac{1}{N} F_1(z_i)Z^{cov}_{g=1} + \dots\),\]
where $F^{(s)}_g(z_i)$ is the exponential of the Liouville action for the genus-$g$ covering map $z(t)$. It is important to note that these functions do not depend on $\mathcal{M}$. Consider the two-point function

\[ \<\s_{[n]}(0) \s_{[n]}(z)\>_{conn} =\frac{1}{(Z_1)^n} \( F^{(s)}_0(z_i) Z^{cov}_{g=0}+F^{(s)}_1(z_i)Z^{cov}_{g=1} + \dots\).\]
In this case $F^{(2)}_0(z_i)= f (n) \; z^{-2 \D_n} $. Accordingly, the renormalization of the twist operators is 

\[ \s_{[n]} \rightarrow \(\sqrt{\frac{(Z_1)^n}{ f_0 Z^{cov}_{g=0}}}+ \mathcal{O} \(\frac{1}{N}\) \) \s_{[n]}. \label{univer}\]
As a consequence, the $s-$point function of the renormalized operators is 

\[ \<\s_{[n_1]}(z_1) \dots  \s_{[n_s]}(z_s)\>_{conn}= \frac{(Z_1)^{\frac{1}{2} \sum_i n_i -\x}}{\sqrt{f(n_1)\dots f(n_s)} (Z^{cov}_{g=0} )^{\frac{s}{2} -1}} \(   F^{(s)}_{g=0}(z_i) +\mathcal{O} \(\frac{1}{N}\) \). \nonumber\]

Both $Z_1$ and $Z^{cov}_{g=0}$ are the partition functions on the sphere of one copy of our seed CFT. We want to argue that the leading term in this expression is independent of the details of the target space $\mathcal{M}$.  This depedence enters only in $Z_1$ and $Z^{cov}_{g=0}$ . In order to evaluate them we need to regularize them. Consider a regulator $\d$ in the base space and a $\d'$ in the covering space. Under the rescaling $z\rightarrow \l z$ the zero-genus partition function transforms as $Z_{1} \rightarrow \l^{c/3} Z_{1} $.\footnote{Because of the Weyl anomaly, under a coordinate transformation the partition function changes like $Z \rightarrow e^{S_L} Z $. Using that $\int \sqrt{-g} R= 8\pi$ for a sphere, when $g=0$ we get $S_L=\frac{c}{3} \ln\d$.}  Thus, we have that $Z_{1}=q \;\d^{\ c/3}$, for some constant $q$ that does depend on $\mathcal{M}$. Similarly, we have that $Z^{cov}_{g=0}=q \;\d'^{\ c/3}$. We see now that the $q$-dependence is $q^g$, and therefore disappears for $g=0$. This shows that correlation function of twist operators are universal. Moreover, one can prove that also the regulators cancel \cite{Lunin:2000yv} in the final expression (note that the functions $f_0(n)$ as well as $F^{(s)}_g (z)$ depend on the regulators).

We now show that the universality and the large-$N$ behavior of correlation functions constrain the OPE of two twist operators. Consider the OPE of $\s_{[n_1]}$ and $\s_{[n_2]}$. From universality, we see that to leading order in $1/N$, the OPE should not depend on the target space $\mathcal{M}$. Correlators involving $\partial X$ have to respect the global properties of $\mathcal{M}$ and depend on its parameters. This implies that to leading order in $1/N$, in the OPE of two twist operators only other twist operators and their orbifold descendants appear, and not operators like those in (\ref{xs}). Combining this fact with the large-$N$ behavior in (\ref{C coeff}), we see that the relevant terms for a four-point function are

\[ \s_{[n_1]} \s_{[n_2]}= \mathbbm{1} \d_{n_1n_2} + :\s_{[n_1]} \s_{[n_2]}: + \frac{1}{\sqrt{N}} \s_{[m]}+ \frac{1}{N}:\s_{[n_3]} \s_{[n_4]}:+ \dots \label{OPE}\]

On the right-hand side, only operators with integer spin can appear, since the left-hand side is invariant under a $2\pi$ rotation around the origin. Let's now consider restrictions coming from group theory. From simple $S_N$ combinatorics (see page 22 in \cite{Lunin:2000yv}) we see that when we combine two single cycles, the possible single cycles that can appear on the right-hand side for $g=0$ are 

\[m=n_1+n_2-2k+1, \quad  1\leq k \leq min\{ n_1,n_2\}. \] 

The double trace operators in (\ref{OPE}) are also restricted\footnote{We thank S. Razamat for this comment.}. From the Riemann-Horowitz formula we have that
\[\frac{1}{2} \sum_j (n_j-1) -\x +1 \geq 0.\]
Since a double trace operator corresponds to a conjugacy class of the permutation group, its two cycles cannot have common elements and therefore 

\[\x\geq n_3+n_4.\]
As a result, we see that in order for the three-point function $\<\s_{[n_1]} \s_{[n_2]}:\s_{[n_3]} \s_{[n_4]}: \>$ to be non-zero, we must have

\[ 0 \leq n_1+n_2-n_3-n_4. \]

For instance, for the cases $n_1=n_2=n_3=2$ and $n_4=4$ this condition is not satisfied and therefore

\[ \< \s_{2}(z_1)\s_{2}(z_2) \s_{(2)(4) }(z_3)\>=0 . \label{2-4}\] 
Hence, we see that in right-hand side of (\ref{OPE}) there is actually a finite number of twisted sectors that appear.

Let us conclude by noting the well-known fact that Virasoro blocks in the large central large limit are reduced to $SL(2,\mathbbm{C})$ blocks \cite{Zamolodchikov:1985ie,Zamolodchikov:1987}, and all the higher Virasoro descendants give subleading contributions. Hence, the Virasoro primaries in (\ref{OPE}) will contribute only an $SL(2,\mathbbm{C})$ block in the large $N$-limit.

\subsection{The zero-genus covering map \label{sec:map}}

In this section, we present a way to construct the zero-genus covering map for a generic four-point function. The covering map for the two and three-point functions was determined in \cite{Lunin:2000yv}. For the two-point function 

\[ \< \s_n(0) \s_n (w)\>, \]
we can use the $SL(2,\mathbbm{C})$ invariance in the covering space to map the two opearator insertions at $t=0$ and $1$, and the map is just 

\[z(t)=a \frac{t^n}{t^n-(t-1)^n}. \]
For the three-point function
\[ \<\s_{n_1}(0) \s_{n_2}(w)  \s_{n_3}(\infty)  \> ,\]
the map is
\[ z(t)=a t^{n_1}  \frac{P_{d_1-n_1}^{(n_1,-d_1-d_2+n_1-1)}(1-2 t)}{P_{d_2}^{(-n_1,-d_1-d_2+n_1-1)}(1-2 t)},\]
where $P^{(a,b)}_n$ are the Legendre polynomials, and
\[d_2=d_1-n_3, \quad d_1=\frac{1}{2}(n_1+n_2+n_3-1 ).\]

For the four-point function, things are much more complicated. In \cite{Pakman:2009zz}, the problem of constructing the map for this case was reduced to solving a second-order differential equation, known as the Heun's equation. However, there is no closed-form general solution for this equation besides some very special cases. In the rest of this section we give a more concrete way to construct the map and we reduce the problem to solving an algebraic equation.

Let us first state the general requirements that such a map has to satisfy. We can use the $SL(2,\mathbbm{C})$ invariance to fix the position of three operators

\[ \<\s_{[n_1]}(0) \s_{[n_2]}(1)  \s_{[n_3]}(w)\s_{[n_4]}(\infty)  \>. \]
We can also use the $SL(2,\mathbbm{C})$ invariance in the covering space to fix $t_1=0$, $t_2=1$ and $t_4=\infty$. The position of the renaming operator will be at some position $t=x$, which is a function of $w$, and will be determined bellow. The map $z(t)$ is a holomorphic function with the following branching behavior

\begin{align}
& \lim_{t \rightarrow 0} z(t) \sim a_1 t^{n_1} \\
& \lim_{t \rightarrow 1} z(t) \sim 1+a_2 (t-1)^{n_2} \\
& \lim_{t \rightarrow x} z(t) \sim w+a_3 t^{n_3} \\
& \lim_{t \rightarrow \infty} z(t) \sim a_4 t^{n_4}. \label{map requirements}
\end{align}
Now let us note that

\[z'(t)=M\; t^{n_1-1}(t-1)^{n_2-1}(t-x)^{n_3-1}, \label{z'} \]
describes a map with $n_4=n_1+n_2+n_3-2$. The next observation is that even if $n_4$ is not equal to $n_1+n_2+n_3-2$, it can always be chosen so that it is smaller than this. Then, the problem of determining the map is to find a way to lower $n_4$ in (\ref{z'}). Let's assume that 

\[n_4=n_1+n_2+n_3-2-2k,\]
for some integer $k$\footnote{The Reimann-Hurwitz formula implies that $n_4-(n_1+n_2+n_3-2)$ is an even integer.}. Then, the above map can be modified to 

\[z'(t)= M \; \frac{t^{n_1-1}(t-1)^{n_2-1}(t-x)^{n_3-1}}{(t-l_1)^2(t-l_2)^2\dots (t-l_k)^2}, \label{map}\]
where the constants $l_i$ are such that $z'(t)$ does not have simple poles at $t=l_i$, which is required so that we don't get logarithms after integration. This requirement can be expressed as

\[Res_{t=l_i} z'(t)=0, \quad \forall i=1,\dots,k.\]
Plugging formula (\ref{map}) into this equation, leads to the following set of algebraic relations for the $l_i$'s

\[\frac{n_1-1}{l_i}+\frac{n_2-1}{l_i-1}+\frac{n_3-1}{l_i-x}=\sum_{j \neq i}^k \frac{2}{l_i-l_j}, \label{li's}\]
which resemble the scattering equations \cite{Cachazo:2013iaa,Cachazo:2013gna}. This set of equations have $k!$ solutions and for each such solution we have $\x$ different functions $t(z)$. These functions are the maps for the ramified maps from a sphere to a sphere, and in the math literature, the numbers of such maps are known as the Hurwitz numbers \cite{lando2003graphs}. Finally, the constant $M$ in (\ref{map}) can be determined by setting $z(t)=1$, or equivalently by

\[1= \int_0^1 dt \; z'(t). \label{K} \]

To summarize, the steps for determining the map for a given set of  $n_i$'s, are:

\begin{itemize}
\item  Choose $n_4$ so that $n_4= n_1+n_2+n_3-2-2k$ for some $k\in \mathbbm{Z}$.
\item  Find the solutions $l_i^{(a)}$ of (\ref{li's}) modulo permutations.
\item  Determine $M^{(a)}(x)$ from (\ref{K}).
\item  Integrate (\ref{map}) to find the function $z^{(a)}(t)=\int_0^t d\tau \;z'(\tau)$.
\item  Finally determine $x_j^a(w)$ by setting $z^{(a)}(x)=w$. 
\end{itemize}

We will show in the next section that the fourth step it's not necessarily required. In \cite{Arutyunov:1997gt} a simpler-looking covering map was obtained for the $k=1$ case. In our approach, when solving for $l$ (there is only one $l_i$ in this case) in (\ref{li's}), the solution will include square roots since it is a second order polynomial in $l$. However, we can avoid the square root by relaxing the condition that the image of $z=1$ is at $t=1$. In fact we can assume that $z(t_1(x))=1$ for some suitable function $t_1(x)$ such that discriminant of (\ref{li's}) is the square of a rational function.

 We conclude this section by  commenting on the covering map for higher genus surfaces. In addition to the requirements outlined in this section, we also have to satisfy the global properties of the covering space. In this case, the map becomes much more complicated, typically expressed in terms of the Weierstrass function. Examples can be found in \cite{Lunin:2000yv,Atick:1987kd}.

\subsection{General Four-Point Function}

Three-point functions of twist operators were calculated in \cite{Lunin:2000yv} (see equation 6.19). The aim of this section is to revisit the calculation of the four-point function

\[ \<\s_{[n_1]}(0) \s_{[n_2]}(1)  \s_{[n_3]}(w)\s_{[n_4]}(\infty)  \>. \]

Using the path integral method, this four-point function was calculated on the covering sphere in \cite{Pakman:2009zz}. Here, we just quote the result

\[G(x,\bar{x})=D |a_1(x)|^{-\frac{c}{12}\frac{n_1-1}{ n_1}}
|a_2(x)|^{-\frac{c}{12}\frac{n_2-1}{ n_2}}
|a_3(x)|^{-\frac{c}{12}\frac{n_3-1}{ n_3}}
| a_4(x)|^{+\frac{c}{12}\frac{n_4-1}{ n_4}}
\(\prod_{i=1}^{k} |C_i(x)|\)^{-\frac{c}{6}}, \]
where the functions $a_i$ were defined in (\ref{map requirements}). The overall constant is given by

\[D=n_1^{-\frac{n_1+1}{12}} n_2^{-\frac{n_2+1}{12}} n_3^{-\frac{n_3+1}{12}} n_4^{+\frac{n_4+1}{12}},\]
and $C_i$'s determined by the behavior of the map close to $l_i$

\[z \approx \frac{C_i}{t-l_i}.\]
Note that here these functions can be determined from the map (\ref{map}) without integrating it. For each solution of (\ref{li's}) we have

\begin{align}
G^{(a)}(x,\bar{x})=&D'|x|^{-\frac{c}{12}\frac{(n_1+n_3)(n_1-1)(n_3-1)}{n_1 n_3}}|1-x|^{-\frac{c}{12}\frac{(n_2+n_3)(n_2-1)(n_3-1)}{n_2 n_3}} \nn \\
& |M^{(a)}(x)|^{-2(h_1+h_2 +h_3-h_4) } \frac{ \prod_i ^k |l^{(a)}_i|^{\frac{c}{3}-4h_1}|1-l^{(a)}_i|^{\frac{c}{3}-4h_2}|x-l^{(a)}_1|^{\frac{c}{3}-4h_3}}{ \prod_{i<j}^k (l^{(a)}_i-l^{(a)}_j)^\frac{c}{3}}, \label{most general G}
\end{align}
where

\[ M^{(a)}(x)= \frac{1}{\int_0 ^1 t^{n_1-1} (t-1)^{n_2-1}(t-x)^{n_3-1} \prod_i (t-l^{(a)}_i)^{-2}},\]
and 
\[D'= n_1^{-\frac{(n_1-1)^2}{12}} n_2^{-\frac{(n_2-1)^2}{12}} n_3^{-\frac{(n_3-1)^2}{12}} n_4^{+\frac{(n_4-1)^2}{12}}.\]
This completes the calculation of the most general four-point function in the covering space. In order to go back to the $z$-coordinates we have to invert

\[w=\frac{\int_0 ^{x^{(a)}} dt \; t^{n_1-1} (t-1)^{n_2-1}(t-x^{(a)})^{n_3-1} \prod_i (t-l^{(a)}_i)^{-2}}{\int_0 ^1 dt \;t^{n_1-1} (t-1)^{n_2-1}(t-x^{(a)})^{n_3-1} \prod_i (t-l^{(a)}_i)^{-2}}, \label{w}\]
and to obtain the final answer, we have to sum over all solutions $x^{(a)}_j(z)$

\[G(z,\bar{z})=\sum_j \sum_a G(x_j^{(a)}(z),\bar{x}_j^{(a)}(\bar{z})). \]

As an example let us look at the case with $n_4=n_1+n_2+n_3-2$, and therefore no $l_i$'s. We immediately see that (\ref{most general G}) simplifies considerably

\[G(x,\bar{x})=D'|x|^{-\frac{c}{12}\frac{(n_1+n_3)(n_1-1)(n_3-1)}{n_1 n_3}}|1-x|^{-\frac{c}{12}\frac{(n_2+n_3)(n_2-1)(n_3-1)}{n_2 n_3}} |M(x)|^{-2(h_1+h_2 +h_3-h_4) }. \label{G for k=0}\]
with
\[ |M(x)|= \frac{ |x|^{1-n_3}}{(n_1 -1)!(n_2 -1)! \; |_1 \tilde{F}_2(n_1,1-n_3,n_1+n_2,1/x)|}, \label{M for k=0}\]
and $_1 \tilde{F}_2$ the regularized hypergeometric function which in this case is reduced to a polynomial. Equation (\ref{w}) is 

\[ w= x^{n_1} \frac{(n_1-1)!}{(n_2-1)!} \frac{ _1 \tilde{F}_2(n_1,1-n_2,n_1+n_3,x)}{ _1 \tilde{F}_2(n_1,1-n_3,n_1+n_2,1/x)} , \label{w(x)}\]
which is a rational function and can be inverted, at least as a power-series in $x$. For small enough $n_i$ it can be inverted exactly, but the result is lengthy, and we don't include it here.

In this section, we have given a algorithm for calculating four-point functions in the symmetric orbifold CFT with our seed CFT being purely bosonic.  However, these results can be used to calculate correlation functions for more general situations, most importantly sypersymmetric theories. In \cite{Lunin:2001pw} it was showed that supersymmetric four-point function of BPS operators factorize to those we have calculated here times correlation functions of spin fields and R-symmetry currents.

\section{Ward Identities \label{sec:Ward}}

So far, we have reviewed how a general correlation function is defined, and we have presented a recipe on how to calculate four-point functions of four twist operators. In this section, we take one step further, and we study insertion of the fractional Virasoro generators. We later will focus on three-point functions. This will be important for better understanding the anatomy of the four-point functions.

From the inversion formula in (\ref{L}), we see that any insertion of $L_{m/n}$ can be computed, once we know

\[ \< T_I(z) \s_{n_1} (w_1) \dots \s_{n_s} (w_s) \>, \quad I=1,\dots, \x .\]
This quantity is apparently easy to calculate \cite{Pakman:2009ab} and the result can be written in a very compact form. The strategy is to write the energy-momentum tensor as

\[T_I(z)= -\frac{1}{2} \lim_{w \rightarrow z} \( \p X^i_I(z) \p X^i_I(w)+ \frac{c}{(z-w)^2} \), \]
and then lift the $(s+1)$-correlation function to the covering space, where it will be just the two-point function of $\p X$. Since $\p X_I$ is not a gauge invariant operator, it will have an image only in a section of the covering space. Let's denote the covering maps for this four-point function by $z_j(t)$ (collectively for the index $j$ and $a$ of section (\ref{sec:map})). For each such map we will have $\x$ inverse functions $t_{j,I}$, in one-to-one correspondence with $\p X_I$. Hence, for each map $z_j(t)$, we have that

\[ \frac{\< \p X^i_I(z) \p X^i_I(w) \s_{n_1} (w_1) \dots \s_{n_s} (w_s) \>_j}{\<  \s_{n_1} (w_1) \dots \s_{n_s} (w_s) \>_j}= -c \frac{t'_{j,I}(z)t'_{j,I}(w)}{(t_{j,I}(z)-t_{j,I}(w))^2} .\]
After subtracting the singularity, the $w \rightarrow z$ limit is

\[ \frac{\< T_I(z) \s_{n_1} (w_1) \dots \s_{n_s} (w_s) \>_j}{\<  \s_{n_1} (w_1) \dots \s_{n_s} (w_s) \>_j}= \frac{c}{12} \{ t_{I,j},z\} ,\]
where $\{t,z\}$ is the Schwartzian derivative (\ref{Schwartian}). Now, we can calculate the insertion of any fractional generator by

\begin{align}
\frac{\< \(L_{m/n_1} \s_{n_1} (w_1) \) \dots \s_{n_s} (w_s) \>_j}{\<  \s_{n_1} (w_1) \dots \s_{n_s} (w_s) \>_j}&= \oint_{w_1} \frac{dz}{2 \pi i} \sum_{I=1} ^{n_1}  e^{2 \pi i m (I-1)/n_1} (z-w_1)^{m/n_1+1}\frac{c}{12} \{ t_{I,j},z\} \nn \\
&=-\frac{c}{12}  \oint_{t_1} \frac{dt}{2 \pi i}  (z_j(t)-w_1)^{m/n_1+1} \frac{\{ z_j,t\}}{z_j'(t)}.
\end{align}
In the second line we lifted the contour integral in the covering space using the properties of the Schwartzian derivative. After we have done this for each $j$, the full correlation function will be the sum of all these terms. Note that, regardless the details of the map, the right-hand side is zero for $m >0$, as it should. 

An important observation is that if we choose $m=-n_1$, then we get a differential equation for the $j$-term of the correlation function. Thus, instead of using the path integral method, which requires a regularization prescription, one can use this differential equation to calculate the correlation function up to a constant. In the literature, this is known as the {\it energy-momentum tensor} method, and it was applied in the case of the four-point function in \cite{Pakman:2009ab}.

The generalization to the case of many $L$'s is now straightforward. We just need to perform all possible Wick contractions of $\p X$ in the covering space, and then take the appropriate limits to get the correlation function with some energy-momentum tensors inserted. For example, for two insertions one gets

\[ \frac{\< T_I(z_1) T_J(z_2) \s_{n_1} (w_1) \dots \s_{n_s} (w_s) \>_j}{\<  \s_{n_1} (w_1) \dots \s_{n_s} (w_s) \>_j}=  \{z_1\}_{j,I} \{z_2\}_{j,J} +\frac{1}{2}  (z_1,z_2)^2_{j,IJ},\]
where we have defined

\[ \{z\}_{j,I}=-  \frac{c}{12}\{t_{j,I}(z),z\},\]
\[ (z_1,z_2)_{j,IJ} =- c\frac{t'_{j,I}(z_1)t'_{j,J}(z_2)}{(t_{j,I}(z_1)-t_{j,J}(z_2))^2}. \]
From this, one can easily calculate the insertion of two $L$'s by evaluating two contour integrals. In practice, in the case of many insertions one has to consider all the possible combinations of the above quantities.

Let us mention that alternatively one could use the usual contour deformation techniques, but this procedure is more involved. The reason is that one cannot start by lifting a $z$-contour integral to the covering space and then deform it, because the $z^{m/n +1}$ introduces a branch cut in the covering space and the contour integral cannot be deformed. Instead, one has to start with a contour integral in the covering space, deform it, and then reduce it to the $z$-space. In this way one does not get just one Virasoro generator but a sum of them, and then has to solve recursively for each of them.

\section{An Example}

Let us now look at the simplest example of a four-point function, namely

\[ \< \s_{[2]}(0)\s_{[2]}(w)\s_{[2]}(1)\s_{[4]} (\infty)\>.\]
with central charge $c=1$. In this case $n_4=n_1+n_2+n_3-2$, and from equations (\ref{G for k=0}) and (\ref{M for k=0}), the four-point function in the covering space is

\[G(x,\bar{x})= \frac{2^\frac{5}{24}}{3^\frac{1}{16}} \frac{|1-2x|^\frac{1}{16}}{ |x(1-x)|^\frac{1}{12}}.\]
Also from (\ref{w(x)})

\[w=x^3 \frac{2-x}{2x-1}. \]

Looking at the $w\rightarrow 0$ limit, the relevant sectors in the OPE of the first two operators, are just

\[ \s_{[2]}\s_{[2]}= \mathbbm{1} \d_{n_1n_2} + :\s_{[2]} \s_{[2]}: + \frac{1}{\sqrt{N}} \s_{[3]}. \]
Because of (\ref{2-4}), the sector $\s_{[2][4]}$ does not appear, and all the other multi-cycle operators give subleading contributions. Of course, in each sector we also have the excited twist operators. We can now invert for $x(w)$, and by expanding for small $w$ we can decompose $G(w,\bar{w})$ into  $SL(2,\mathbbm{C})$ conformal blocks $G_{h,\bar{h}}(w,\bar{w})$. The first few blocks are

\begin{align}
G(u,\bar{u})= G_{\frac{1}{9},\frac{1}{9}}+G_{\frac{1}{9}+\frac{2}{3},\frac{1}{9}+\frac{2}{3}}+\(G_{\frac{1}{9}+2,\frac{1}{9}}+G_{\frac{1}{9},\frac{1}{9}+2}\)+\dots  G_{\frac{1}{8},\frac{1}{8}}+\( G_{\frac{1}{8}+2,\frac{1}{8}}+ G_{\frac{1}{8},\frac{1}{8}+2}\) +\dots \nn
\end{align}
which correspond to 

\[\s_{[3]} + L_{-\frac{2}{3}} \bar{L}_{-\frac{2}{3}} \s_{[3]} + \(L_{-\frac{2}{3}}\)^3\s_{[3]} +\(\bar{L}_{-\frac{2}{3}}\)^3\s_{[3]}+\dots + \s_{[2]} \s_{[2]}+ \s_{[2]} \p\p \s_{[2]}+\s_{[2]} \bar{\p}\bar{\p} \s_{[2]} \dots \nn  \]
We can further verify the relative OPE coefficients using the discussion of the previous section. To make contact with recent developments, we can also state the expectations for a Mellin representation \cite{Mack:2009gy,Mack:2009mi,Penedones:2010ue} of such a correlator. Using the conventions of \cite{Rastelli:2017udc}, we expect the following form

\[ G(w,\bar{w})=\int_\g ds dt |w|^{s-\frac{1}{4}} |1-w|^{t-\frac{1}{4}} \mathcal{M}(s,t) \; \Gamma(\frac{1}{8}-\frac{s}{2}) \Gamma(\frac{1}{8}-\frac{t}{2}) \Gamma(\frac{1}{8}-\frac{u}{2}), \]
where $u+s+t=\frac{11}{16}$, and the contour $\g$ is such that the poles in $s$ and $t$ are on one side. Some remarks are in order. First of all, we expect just three gamma functions on the right-hand side, since the double trace operator $\s_2 \s_4$ does not contribute. Furthermore, we can predict the poles of the Mellin amplitude $\mathcal{M}$ from the twist (scaling dimensions minus spin) of the $SL(2,\mathbbm{C})$ primaries in the family of $\s_3$. Hence, we expect poles only at

\[s,t=\frac{2}{9}+\frac{4}{3}k+2m , \quad k,m \in \mathbbm{Z}.  \]

Given that we have an infinite number of primary operators, we expect an infinite number of poles.

One last comment is the application of the invesion formula for the OPE coefficients \cite{Caron-Huot:2017vep,Simmons-Duffin:2017nub}. The inversion formula for an arbitrary four-point function in Euclidean space, is

\[C(\Delta ,J)=N_{h ,\bar{h}}\int d^2w \; |1-w|^{\frac{1}{2} \left(\Delta _2+\Delta _3-\Delta _1-\Delta _4\right)}|w|^{\frac{1}{2} \left(\Delta _1+\Delta _2\right)-2}F_{h ,\bar{h}}\left(w,\bar{w}\right)G\left(w,\bar{w}\right). \nn \]
where $F_{h,\bar{h}}$ is the conformal partial wave in two dimensions \cite{Osborn:2012vt}, and the constant in front of the expression is given by

\[N_{h ,\bar{h}}=\frac{1}{2\pi} \frac{\kappa_{2 h}}{\kappa_{2-2 \bar{h}}},\quad \kappa_{\b}=\frac{\Gamma(\frac{\b+\Delta_2-\Delta_1}{2}) \Gamma(\frac{\b-\Delta_2+\Delta_1}{2}) \Gamma(\frac{\b+\Delta_3-\Delta_4}{2}) \Gamma(\frac{\b-\Delta_3+\Delta_4}{2}) }{ \Gamma(\b-1)\Gamma(\b)} .\]
The inversion formula can be easily lifted to the covering space

\begin{align}
C(\Delta ,J)=N_{h ,\bar{h}}\int d^2x &  \; |w'(x)|^2|1-w(x)|^{\frac{1}{2} \left(\Delta _2+\Delta _3-\Delta _1-\Delta _4\right)}|w(x)|^{\frac{1}{2} \left(\Delta _1+\Delta _2\right)-2} \nn\\
&\times F_{h ,\bar{h}}\left(w(x),\bar{w}(\bar{x})\right)G\left(x,\bar{x}\right) .
\end{align}
However, the result is complications. First, even if one can use the integral representation of the partial wave, the resulting integral is very complicated to be solved analytically. Secondly, the integral in not finite and one needs to subtract the non-normalizable contributions \cite{Simmons-Duffin:2017nub}, namely all the contributions from operators with scaling dimensions less than one. Thus, first one needs to subtract these contributions and then lift it to the covering space.

\section*{Acknowledgments}

I am grateful to Leonardo Rastelli for suggesting the project, as well as providing guidance throughout all the stages of this work. I would also like to thank  Samir Mathur, Shlomo Razamat and Martin Rocek for useful discussions, and Anna-Maria Taki for editorial help. My work is supported by the NSF Grant PHY-1620628.

\section*{Note Added} After this work was completed, reference \cite{Burrington:2018upk} appeared to overlap with ours to some extent. The authors of that paper discuss some topics that we consider here, like the fractional Virasoro generators and the large $N$-limit of correlation function of twist operators.

\bibliographystyle{JHEP}

\bibliography{refs}

\end{document}